\documentstyle[epsfig]{aa} 

\def\kms{km~s$^{-1}$} 
\def\cc{$^{\circ}$} 
\def\cds{cd$^{-1}$\,} 
\def\cd{cd$^{-1}$}
\def\I{\'\i} 

\begin{document} 
\title{Preparing the COROT space mission: incidence and characterisation of
pulsation in the Lower Instability Strip\thanks{Based on 
observations collected at S.Pedro Mart\I r, Sierra Nevada, La Silla,
Haute--Provence, South African
and Roque de Los Muchachos observatories.}
} 
\author{E.~Poretti\inst{1}, R.~Garrido\inst{2}, P.J.~Amado\inst{2}, 
K.~Uytterhoeven\inst{3}, G.~Handler\inst{4,5}, 
 R.~Alonso\inst{6}, S.~Mart\I n\inst{1}, C.~ Aerts\inst{3}, 
 C.~Catala\inst{7}, M.J.~Goupil\inst{7}, E.~Michel\inst{7}, 
L.~Mantegazza\inst{1}, P.~Mathias\inst{8}, M.L.~Pretorius\inst{9}, 
J.A.~Belmonte\inst{6}, A.~Claret\inst{2}, 
 E.~Rodr\I guez\inst{2}, J.C.~Suarez\inst{2,7}, F.F.~Vuthela\inst{4,10},
W.W.~Weiss\inst{5},
D. Ballereau\inst{11}, J.C.~Bouret\inst{12}, S.~Charpinet\inst{13}, 
T.~Hua\inst{12}, T.~L\"uftinger\inst{5}, N.~Nesvacil\inst{5}, 
C.~Van't~Veer-Menneret\inst{11}
} 
\institute { 
INAF-Osservatorio Astronomico di Brera, Via Bianchi 46,
I-23807 Merate, Italy 
\and Instituto de Astrof\I sica de Andaluc\I a, C.S.I.C., Apdo. 3004, 18080 Granada,
Spain
\and Instituut voor Sterrenkunde, Katholieke Universiteit Leuven, Celestijnenlaan 
200 B, B-3001
Leuven, Belgium
\and South African Astronomical Observatory, P.O. Box 9, Observatory
7935, South Africa
\and Institut f\"ur Astronomie, Universit\"at Wien, T\"urkenschanzstrasse
17, 1180 Wien, Austria 
\and Instituto de Astrof\I sica de Canarias, C/ V\I a L\'actea s/n, 38200 La 
Laguna, 
Tenerife, Spain
\and Observatoire de Paris, LESIA, FRE 2461, F-92195, Meudon, France
\and Observatoire de la C\^ote d'Azur, UMR 6528, BP 4229, F-06304 Nice Cedex, 
France 
\and Dept. of Astronomy, University of Cape Town, Rondebosch 7700, South Africa 
\and Dept. of Physics, University of the North-West, Private Bag X2046,
Mmabatho 2735, South Africa
\and Observatoire de Paris, GEPI, F-92195, Meudon, France 
\and Laboratoire d'Astrophysique de Marseille, France
\and Laboratoire d'Astrophysique, Observatoire Midi-Pyr\'en\'ees, 14 avenue E.
Belin, 
31400 Toulouse, France 
} 

\offprints{E. Poretti \\ 
 \email{poretti@merate.mi.astro.it}} 
\date{Received date; Accepted Date}

\abstract{
By pursuing the goal to find new variables in the COROT field--of--view we 
characterised a sample of stars located in the lower part of the 
instability strip. Our sample is composed of stars belonging to the disk 
population in the solar neighbourhood. We found that 23\% of the stars
display multiperiodic light variability up to few mmag of amplitude, i.e., 
easily detectable in a single night of photometry. $uvby\beta$ photometry
fixed most of the variables in the middle of the instability strip and
high--resolution spectroscopy established that they have $v\sin i >$100 
\kms. An analysis of the Rodr\I guez \& Breger (2001) sample ($\delta$
Sct stars in the whole Galaxy) shows slightly different features, i.e., 
most $\delta$ Sct stars have a 0.05--mag redder $(b-y)_0$ index and 
lower $v\sin i$ values. Additional investigation in the open cluster
NGC~6633 confirms the same incidence of variability, i.e., around 20\%. 
The wide variety of pulsational 
behaviours of $\delta$ Sct stars (including unusual objects such as a 
variable beyond the blue edge 
or a rapidly rotating high--amplitude pulsator) makes them  
very powerful asteroseismic tools to be used by COROT. Being quite 
common among bright stars, $\delta$ Sct stars are suitable targets for
optical observations from space. 
} 

\authorrunning{Poretti et al.}
\titlerunning{$\delta$ Sct stars for COROT}
\maketitle

\section{Introduction} 

The preparation of new asteroseismic space missions requires  a
considerable amount of related 
theoretical and observational work. In the domain of $\delta$ Sct and
related stars the determination of the pulsational spectra has so far
been performed by means of intensive, single-- or multi--site, campaigns, both
photometric and spectroscopic. The attention of a wide community of
specialists has been mostly focused on some well--defined case studies (FG
Vir, XX Pyx, 4~CVn, $\theta^2$ Tau, BI CMi, 44~Tau; see Poretti 2000 and
Breger 2000 for reviews). The scientific plan of the European mission
COROT (COnvection, ROtation and planetary Transits; Baglin et al. 2002)
has a slightly different strategy, as the satellite will monitor selected
targets located in two fields centered at $\alpha=18^h50^m$,
$\delta=0$\cc\, (i.e., in the direction of the Galactic Center), and
$\alpha=6^h50^m$, $\delta=0$\cc\, (Anticenter direction), each having a
semi-aperture of 10\cc. None of the case studies mentioned above is
included in these fields.

Consequently suitable candidate target stars have to be searched for. 
They should be chosen so as to provide a good trade--off between being 
challenging for aspects of theoretical modelling and having suitable observable 
features. 
The $\delta$ Sct class 
covers both the early main-sequence evolutionary stage, when the star is 
burning hydrogen in the core, and the following one, when it 
leaves the Terminal--Age Main Sequence (TAMS) 
burning hydrogen in a shell. These two different 
stages correspond to different types of structures and their study offers
different insights into the physics of the stellar interiors
(Breger \& Pamyatnykh 1998). 
Therefore, both types of variables deserve interest. 
However, the primary objectives of COROT are highly focused on 
core overshooting processes and on transport of angular momentum and
chemical species. These processes are crucial in the main--sequence stage
for intermediate-- and high--mass stars. Therefore, we put a strong 
priority in searching for new $\delta$ Sct stars close to the ZAMS.

Evolved  $\delta$ Sct variables pose considerable problems  in terms
of analysis which limits their use in tests of theory.
It is known that evolved
models of $\delta$ Sct stars have such a dense frequency spectrum that
matching theoretical and observed frequencies might be a hopeless task
without further input.  As an example, one can consider the prototype of the class,
$\delta$ Sct itself, which is included in one of the fields accessible to COROT.
Templeton et al. (1997) constructed evolution and pulsation models that
match the observed spectral type, luminosity and the identified radial
mode frequency of $\delta$ Sct. Accounting for rotational splitting, there
are 275 possible $\ell\le2$ modes in a 4 \cds range (Guzik et al. 2000).  
Therefore, asteroseismic inferences for such a star will be complicated
even for long runs from space.

Rotation is another aspect to have in mind when evaluating the scientific
perspectives of a given target. Fast rotation considerably complicates the
modelling of the oscillations. By extending the perturbation theory
of the influence of rotation on stellar oscillations to third
order (Soufi et al. 1998), asteroseismic inference from oscillations can
include relatively high rotation rates. For the mass range of $\delta$ Sct
variables this means that stars with equatorial rotational velocities up
to 100~\kms can be modelled, but beyond that the application of perturbation
theory remains questionable. Ideally, we would like to scan different
rotation rates if being sure to have reasonably slow rotators among the
targets. Finally, stars showing variability at the level of a few mmag
(at least) should be preferred as they should guarantee the possibility to
complement photometry from space with high--resolution spectroscopy from
ground.

We report here on how we tackled the problem of the identification of 
suitable targets by means of a theoretical selection followed by an 
observational one.

\section{Observations and data reduction} 

The search for new $\delta$ Sct stars in the COROT field of view has been
performed in two observational steps. The first is related to the
ground--based activity build--up to determine the physical parameters of 
all the stars brighter than $V$=8.0 which are included in the COROT 
accessible fields. This requires that all these stars should be observed at
least once in the Str\"omgren system and at least once with 
high--resolution spectroscopy. In the Center direction this program
was completed
well before the summer of 2002. Therefore we know the $uvby\beta$
indices and the $v\sin i$ values for most of our targets. 
Photometric 
observations were carried out at Sierra Nevada Observatory (automatic 
six--channel spectrophotometer at the 90--cm telescope), spectroscopic
ones at Haute--Provence Observatory ({\sc elodie} instrument at the 
193--cm telescope) and at La Silla Observatory ({\sc feros} instrument at
the 152--cm telescope). 
The reduction of the photometric data and their transformation into the 
standard system has been done following the procedures described in 
Olsen (1993) and references therein. The results of these procedures 
applied to our dataset will be presented in a future work (Amado et al., 
in preparation).
The $v\sin i$ determinations have been
performed taking into account instrumental broadening and 
limb--darkening effects; uncertainties are of the order of 5--6 \kms. 

The second observational task is specific to our program. Once the 
potential targets have been selected, dedicated observing programs have 
been carried out at Sierra Nevada Observatory,  at S. Pedro Mart\I r 
Observatory ($uvby$ photometry at
the 152--cm telescope -- twin photometer of the one at Sierra Nevada), at
South African Astronomical Observatory (50 and 75--cm telescopes with $v$
and $y$ filters) and at the Mercator telescope (120--cm telescope on
Canary Islands, Geneva photometric system). Unfortunately, it has not been  
possible to undertake simultaneous two--site campaigns. We decided to 
monitor 4--5 stars each night, changing the group of stars every night. Of 
course, such a strategy cannot be considered as totally conclusive about 
the variability of a specific star. However, our main goal was to
scrutinise the whole sample, and to make clear variability detections at 
the 0.005--mag level. 

\section{The identification of potential targets} 

\begin{figure}[] 
\resizebox{\hsize}{!}{\includegraphics{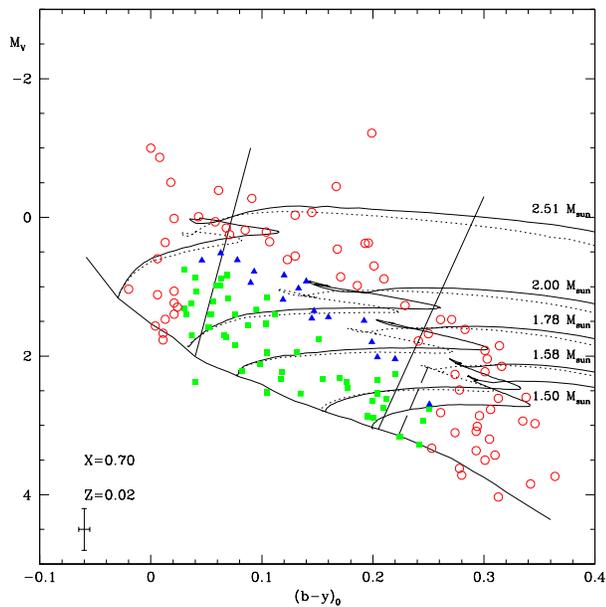}}

\caption{{\sc hipparcos} unreddened $M_V$ against our dereddened $(b-y)_0$ 
colour indices for potential COROT targets in the Center direction. 
Dotted and solid lines indicate evolutionary tracks for 
$d_{\rm over}$=0.1 and $d_{\rm over}$=0.2, respectively.
Solid squares represent stars surely unevolved, independently from
overshooting influence.  Solid triangles represent stars
whose evolutionary status depends on the overshooting importance. 
Open circles represent stars too advanced on evolutionary
tracks or too far outside the instability strip. The borders of the 
$\delta$ Sct instability strip and the edge of the $\gamma$ Dor domain
are also indicated. 
} \label{cmd}
\end{figure}

The $uvby\beta$ photometry performed at OSN allowed us to construct a 
colour--magnitude diagram (CMD), which will be one of our main tools to 
identify the best targets (Fig.~\ref{cmd}). We considered some B stars and 
all the stars belonging to the A and F spectral types; this sample 
comprises 138 stars. $uvby\beta$ colour indices were dereddened following the 
procedure described by Philip et al. (1976); only in 18\% of the cases
were the corrections larger than 0.05 mag. 

The apparent magnitudes were dereddened by using the relation given by Crawford 
\& Mandwewala (1976). 
The {\sc hipparcos} parallaxes were used to determine
the absolute magnitude $M_V$. 
In the cases where the {\sc hipparcos}
parallaxes were not available, $M_V$ values were derived from the
Str\"omgren indices by using standard photometric calibrations.

We also need to  define  ``lower part of the 
instability strip". The main sequence has been taken from Philip \& Egret
(1980) and the $\delta$ Sct instability strip borders from Rodr\I guez \&
Breger (2001). The red border of the $\gamma$ Dor instability strip has 
been taken from Handler \& Shobbrook (2002); the blue border is well
inside the $\delta$ Sct instability strip. We also added evolutionary 
tracks, considering five values for the mass in the range from 1.50 to
2.51 M$_{\sun}$. We calculated two sets of models for two different 
typical overshooting extension distances, i.e., $d_{\rm over}$=0.1 
and $d_{\rm over}$=0.2 
(see Claret 1995 for details). At this point the CMD shown in 
Fig.~\ref{cmd} is constructed.

Our initial sample consisted of 138 stars; 70 stars out of 138 fall 
outside the region of our interest, i.e., they are too far from the red and blue 
borders of the instability strip or too advanced on the evolutionary
tracks of models with $d_{\rm over}$=0.2. However, even considering all the stars 
below the zigzags of $d_{\rm over}$=0.2 tracks, several of them 
could actually be evolved stars, depending on the model of choice, with or 
without overshooting.

To avoid a bias originating from the theory, we divided the stars on the 
basis of the two different models. Stars located close to ZAMS are 
considered unevolved objects and therefore high--priority targets for 
the COROT preparation program. On the other hand, the stars approaching 
the zigzags of 
$d_{\rm over}$=0.2 tracks could be evolved ones if the overshooting is not 
as effective as supposed. These stars are low--priority objects for the 
COROT program. As can be seen in Fig.~\ref{cmd}, limits have been
a little relaxed to take into account possible uncertainties in the data,
in the dereddening relation, 
in the calibrations, in the border definitions and so on. Overall average
errors of the data and  the hydrogen and metal contents of the models
are shown in the lower left corner.

Among the 68 selected stars, three variables are known already: HD 183324
(a $\lambda$ Boo star; Paunzen et al. 2002), HD 167858 and HD 175537 (two
$\gamma$ Dor stars; Handler 1999). Moreover, HD 182475 and 
HD 177702 were good 
candidates to be $\delta$ Sct variables on the basis of previous surveys 
(Hildebrandt 1992 and Handler 2002, respectively): we kept them in our
sample for confirmation. Therefore, we have 65 stars that are good 
theoretical candidate $\gamma$ Dor or $\delta$ Sct pulsators.

\section{The detection of variability}
\begin{table}
\caption{New and known variable stars located in the COROT
accessible field, Center direction.} 
\begin{tabular} { r c rl rr } 
\hline 
\multicolumn{1}{c}{Star} & 
\multicolumn{1}{c}{} & 
\multicolumn{1}{c}{$V$} & \multicolumn{1}{c}{Sp.} & \multicolumn{1}{c}{$v \sin i$} 

& \multicolumn{1}{c}{$v$ ampl.}\\
\multicolumn{1}{c}{} & 
\multicolumn{1}{c}{} & 
\multicolumn{1}{c}{} & \multicolumn{1}{c}{} & \multicolumn{1}{c}{[\kms]} 
& \multicolumn{1}{c}{[mmag]}\\
\hline 
\noalign{\smallskip}
\noalign{New unevolved $\delta$ Sct stars} 
\noalign{\smallskip}
HD 170699 & & 6.95 & A3 & $>$200 & 30 \\
HD 170782 & & 7.81 & A2 & 198 & 6 \\ 
HD 171234 & & 7.91 & A4 & 162 & 7 \\ 
HD 174966 & & 7.70 & A3 & 125 & 40 \\ 
HD 176112 & & 7.98 & F0 & 117 & 30 \\ 
HD 181555 & & 7.98 & A5 & 170 & 30 \\ 
\noalign{\smallskip}
\noalign{New evolved $\delta$ Sct stars}
\noalign{\smallskip}
HD 174532 & & 6.91 & A2 & 32 & 25\\
HD 177064 & & 7.74 & A2 & 180 & 50\\ 
HD 177702 & & 7.30 & F0 & $>$200 & 150 \\ 
HD 182475 & & 6.61 & A9V & 139& 35\\ 
\noalign{\smallskip}
\noalign{Known unevolved $\delta$ Sct star}
\noalign{\smallskip}
 HD 183324 & & 5.79 & A0V & 98 & \\
\noalign{\smallskip}
\noalign{Known $\gamma$ Dor variables}
\noalign{\smallskip}
HD 167858 & & 6.62 & F2V & 10 &\\
HD 175337 & & 7.39 & F5 & 39 & \\
HD 169577 & & 8.64 & F0 & & \\
\noalign{\smallskip}
\noalign{New very evolved $\delta$ Sct star} 
\noalign{\smallskip}
HD 172588 & & 7.22 & F0II-III & 12 & \\ 
\noalign{\smallskip}
\noalign{Known very evolved $\delta$ Sct stars} 
\noalign{\smallskip}
 \multicolumn{1}{l}{$\delta$ Sct}& & 4.70 & F2IIIp &27 & \\ 
HD 174553 & & 9.35 & F8 & 42 & 130\\ 
\noalign{\smallskip}
\noalign{Suspected $\gamma$ Dor star} 
\noalign{\smallskip}
HD 178596 & & 5.24 & F0III-IV & 77 &15 \\ 
\noalign{\smallskip}
\noalign{Geometrical variables ?}
\noalign{\smallskip}
HD 171802 & & 5.37 & F5III& 13 & 40\\ 
HD 172506 & & 7.96 & F2 & 47 & 20\\
HD 179123 & & 7.40 & A5 & 10 & 10 \\ 
\noalign{\smallskip}
\hline 
\end{tabular}
\label{soi} 
\end{table}

Despite the large number of stars to be observed and the required accuracy 
in the measurements, only a few stars could not be evaluated. Four stars 
have a close companion: they cannot be measured as single stars not only 
from ground, but neither from space, as the defocused images of two close
stars will result in inaccurate photometry. Another star is too bright to
be measured by COROT. For one star we could obtain only a
few measurements suggesting constant brightness.

Therefore, we successfully monitored 59 stars. We discovered at least 13 
stars displaying evident light variations and we confirmed the $\delta$ 
Sct variability of HD~182475 and HD 177702, obtaining more reliable light
curves. All these variables are listed in Table~\ref{soi}.

\subsection{The new $\delta$ Sct stars} 

\begin{figure}[] 
\resizebox{\hsize}{!}{\includegraphics{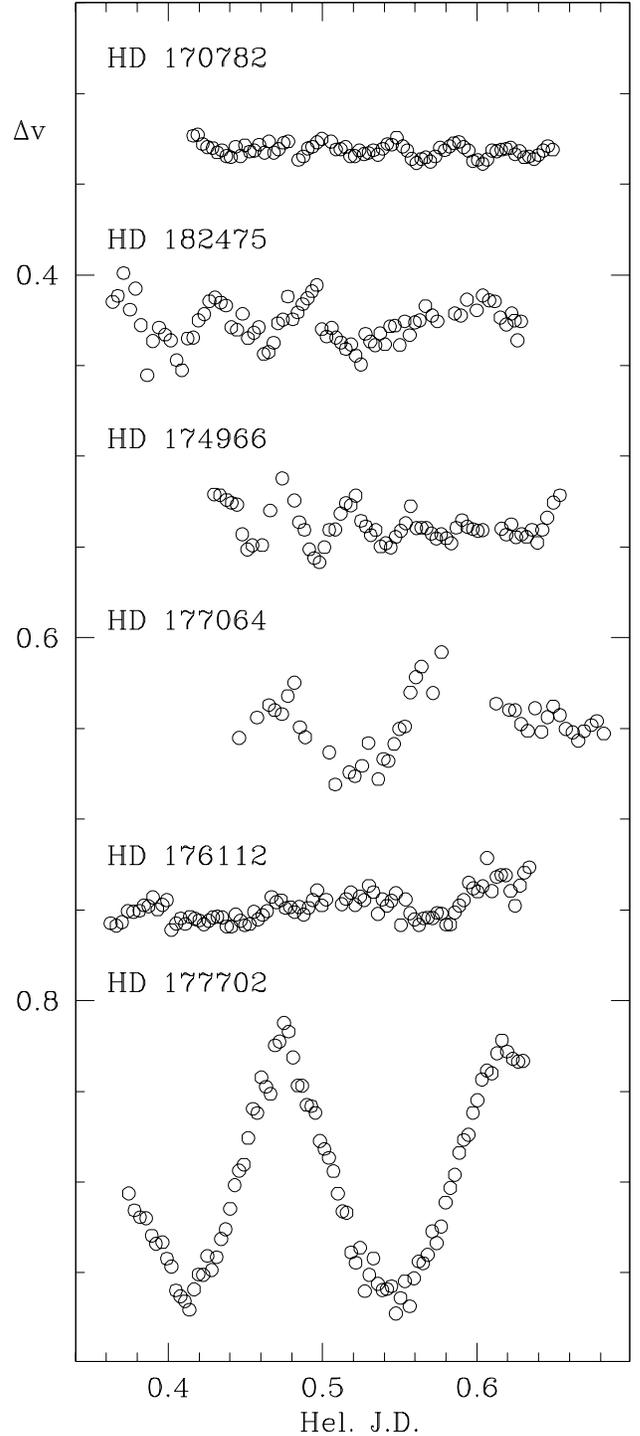}} 
\caption{Light curves of some $\delta$ Sct discovered in the field 
accessible to COROT.}
\label{zoo} 
\end{figure}

The sample of $\delta$ Sct stars discovered in the Center direction is
quite representative of the different types of behaviour known of this class of 
pulsating star (Fig.~\ref{zoo}). 

The simplest variables are HD 170782 and HD 171234, which show almost 
regular light curves. The frequency analysis of the HD 170782 
(Fig.~\ref{zoo}, curve at top) data yields a main term at $f$=25.0 \cd\, and 
the subsequent least--squares fit a full--amplitude of 6.0 mmag in $v$
light, with a rms residual of 2.8 mmag. In the case of HD 171234 we got a
main term at $f$=23.6 \cd, a full--amplitude of 6.8 mmag in $v$ light and
a rms residual of 4.0 mmag. 

The multiperiodic behaviour of HD 177064, HD 170699, HD 174966, HD 182475
and HD 181555 is clearly visible in a single night. For example, during
the second half of the light curve of HD 182475, the rapid variability
disappears and a long cycle takes place (Fig.~\ref{zoo}, second curve from
top). Changes in the total amplitude oscillation are seen in HD 170699,
HD 174966
(Fig.~\ref{zoo}, third curve from top) and HD 177064 (Fig.~\ref{zoo}, 
fourth curve from top). The photometric variability of HD 174966 is 
corroborated by the spectroscopic detection of line profile variations. HD 
181555 shows other features in its light curve, i.e., oscillations are
always visible, but with changing shapes and different time intervals 
between consecutive extrema.

The light curve of HD 176112 (Fig.~\ref{zoo}, fifth curve from top) adds 
more complications, as it suggests the possibility of a long term
variation ($\gamma$ Dor pulsation or geometrical variability) superposed 
to a rapid one ($\delta$ Sct pulsation).
Further 
observations are necessary to characterise this variable star better, 
also considering that {\sc hipparcos} photometry is inconclusive in this respect. 

To conclude the journey along the different behaviours of our $\delta$ Sct 
stars, HD 177702 (Fig.~\ref{zoo}, curve at bottom) provides a good example 
of a high amplitude light curve (0.17 mag in $v$). High--resolution 
spectroscopy allowed us to determine $v\sin i>$200~\kms ,
a value quite unusual for a high--amplitude $\delta$ Sct star. 

\subsection{Stars showing slow variability}

When observing such a large sample, we
should expect some unclear cases, especially measuring each target on one
night only. Indeed, we found five of these unclear cases. 

The light curves of HD 172506 and HD 171802 show a continuous change in 
brightness, up to 0.04 mag in $v$~light. The spectral types of both stars
are compatible with those of $\gamma$ Dor pulsators. However, the
variability is not discernible in the $v-y$ colour curve and therefore
geometrical variability due to binarity seems more probable. We also note
that the high--resolution spectrogram of HD~172506 does not show two sets
of spectral lines or noticeable line profile deformations. Moreover, the 
analysis of the {\sc hipparcos} photometry did not show any evidence for 
variability: the amplitude spectra have peak noise levels of 3 mmag for HD 
171802 and 8 mmag for HD 172506. 
In the case of HD 179123 the $v$--amplitude is only 0.01 mag. The star
can be a $\gamma$ Dor variable ($(b-y)_0$=0.212), but a rotational 
or even a spurious effect are also plausible.

Therefore, the variability of HD 172506, HD 171802 and HD 179123 
needs further confirmation; the three stars are considered
non--pulsating stars (reported as ``Geometrical 
variables~?" in Tab.~\ref{soi}) in the following discussion.

On the other hand, the drift in the HD 178596 data is visible both in the
$v$ and in the $v-y$ curves. Taking also into account that the 
high--resolution spectrograms show line profile deformations
 and that the star is located close to the red border of the
instability strip, the $\gamma$ Dor hypothesis seems preferable to a
geometrical variability (the star is reported as
``Suspected $\gamma$ Dor" in Tab.~\ref{soi}). However, we do not include it in
the list of pulsating variables, waiting for further confirmation; we 
note that the
{\sc hipparcos} photometry is also inconclusive.

Finally, just for one star (HD 173611) the light curve shows a scatter
which did not allow us to select between variability or constancy. 

\subsection{Constant stars} 

Forty--four stars did not display detectable light variation. Of course, 
they can be variables on longer time scales or can have very small 
amplitude, not detectable in our short single--night runs. In the former 
case they would not be of particular interest, in the latter case the 
observational effort to detect variability would require a long and 
intensive survey which we cannot undertake at this stage for such a large
sample. We note that some photometrically constant stars actually show
spectroscopic variability. This is not surprising, as it is well known
that high--degree nonradial pulsation modes can produce cancellation 
effects in photometry, but can still be detectable by spectroscopy. 

The quality of a given night biases the variability detection. As an
example, the standard deviations of the time series on HD 185090, HD
170274 and HD 180086 (obtained on the nights of JD 2452488 and 2452489) 
and on HD 176921 and HD 177011 (obtained on the night of JD 2452467) are 
well above 6.5 mmag, suggesting non--perfect photometric nights. In this 
context, we have to stress that the strategy to observe each group of 
stars just in one night was the only viable one, taking into account the large 
number of stars to be monitored and the limited telescope time available.

\begin{table}
\caption{Stars not showing an evident trace of variability. They are
considered as {\it constant} for our purposes. N is 
the number of measurements; s.d. is the standard deviation. The note ``--" 
indicates stars used once as comparison stars and not re--observed.}
\begin{tabular} { r c rl r  rr} 
\hline 
\multicolumn{1}{c}{Star} & 
\multicolumn{1}{c}{} & 
\multicolumn{1}{c}{$V$} & \multicolumn{1}{c}{Sp.} & \multicolumn{1}{c}{$v \sin i$} 

& \multicolumn{1}{c}{N} &
\multicolumn{1}{c}{s.d.}\\ 
\multicolumn{1}{c}{} & 
\multicolumn{1}{c}{} & 
\multicolumn{1}{c}{} & \multicolumn{1}{c}{} & \multicolumn{1}{c}{[\kms]} 
& \multicolumn{1}{c}{} & \multicolumn{1}{c}{[mmag]}\\
\hline 
\noalign{\smallskip}
\noalign {Unevolved stars} 
\noalign{\smallskip}
HD 166991 & & 6.83 & A2 & 184 & \multicolumn{2}{c}{--} \\ 
HD 167946 & & 7.34 & A0 & 53 & 76 & 3.2\\ 
HD 167968 & & 7.75 & A2 & 199 & 58 & 3.1\\ 
HD 169268 & & 6.36 & F6 & 20 & 52 & 3.9 \\ 
HD 169436 & & 7.71 & F2 & 162 & 59 & 3.8\\ 
HD 170818 & & 7.24 & F2 & 86 & 79 & 3.5\\ 
HD 171149 & & 6.35 & A0Vn& 288 & 62 & 3.2\\
HD 171834 & & 5.44 & F3V & 72 & 69 & 2.8 \\
HD 171836 & & 7.70 & F0 & 60 & \multicolumn{2}{c}{--}\\
HD 173073 & & 7.67 & A0 & 66 & \multicolumn{2}{c}{--} \\ 
HD 173369 & & 7.99 & A2 & 23 & 90 & 5.2\\ 
HD 174162 & & 7.76 & A0 & 156 & 58 & 5.9\\ 
HD 175272 & & 7.44 & F5 & 23 & \multicolumn{2}{c}{--}\\
HD 175543 & & 7.06 & A5V & 12 & 59 & 4.5 \\
HD 176921 & & 8.00 & A2 & $<$10 & 25 & 8.5\\ 
HD 177011 & & 7.20 & A0 & $<$10 & 25 & 7.3 \\
HD 177177 & & 7.82 & A2 & $<$10 & 45 & 6.4\\ 
HD 177178 & & 5.83 & A4V & 176 & 27 & 2.7 \\ 
HD 177332 & & 6.72 & A5m & 90 & \multicolumn{2}{c}{--} \\ 
HD 177552 & & 6.54 & F1V & 41 & 50 & 3.7 \\
HD 177959 & & 7.28 & A3 & 136 & 61 & 3.8 \\
HD 178190 & & 7.11 & A2 & 151 & 57 & 2.7\\ 
HD 178265 & & 7.19 & F0 & 54 & \multicolumn{2}{c}{--} \\ 
HD 178409 & & 7.90 & A0 & 54 & 67 & 4.1\\ 
HD 178857 & & 7.72 & A0 & 58 & 25 & 2.4\\ 
HD 178954 & & 6.83 & A0 & 144 & \multicolumn{2}{c}{--} \\ 
HD 179739 & & 7.90 & A2 & & 31 & 7.0 \\ 
HD 179742 & & 7.66 & F1 & $<$10 & 25 & 2.6\\ 
HD 179892 & & 7.82 & Am & 90 &49 & 5.5\\
HD 179939 & & 7.22 & A3 & 52 & \multicolumn{2}{c}{--}\\
HD 181414 & & 7.07 & A2 & 8 & \multicolumn{2}{c}{--}\\ 
HD 182623 & & 7.82 & A0 & $<$10 &\multicolumn{2}{c}{--}\\ 
HD 183265 & & 7.33 & A0 & & 50 & 4.6\\
HD 185090 & & 7.31 & A5 & & 72 & 6.7\\
\noalign{\smallskip}
\noalign {Evolved stars} 
\noalign{\smallskip}
HD 169310 & & 7.52 & A3 & 134 & 52 & 5.7\\ 
HD 169725 & & 6.85 & A3 & 16 & 52 & 4.2 \\ 
HD 170274 & & 7.86 & F1 & 31 & 65 & 8.3 \\ 
HD 174589 & & 6.08 & F2III & 97 & \multicolumn{2}{c}{--} \\ 
HD 174866 & & 6.37 & A7Vn& 184 & 67 & 5.3 \\ 
HD 175015 & & 7.81 & A0 & 135 & \multicolumn{2}{c}{--}\\ 
HD 175250 & & 7.05 & B9 & & \multicolumn{2}{c}{--}\\ 
HD 175664 & & 7.75 & A0 & 51 & 63 & 5.1\\ 
HD 176074 & & 7.09 & A2 & 18 & \multicolumn{2}{c}{--}\\
HD 180086 & & 6.63 & F0 & 197 & 72 & 6.8 \\
\hline 
\end{tabular}
\label{cst} 
\end{table} 

\section{Impact of the results on general scientific topics and on specific
COROT aspects} 

The results described above will be used to characterize the potential
COROT targets and to build the observational program. Besides that, they
can be used to investigate pulsation in the lower instability strip.

\subsection{The variability in the lower part of the instability strip}

\begin{figure} 
\includegraphics[scale=0.40]{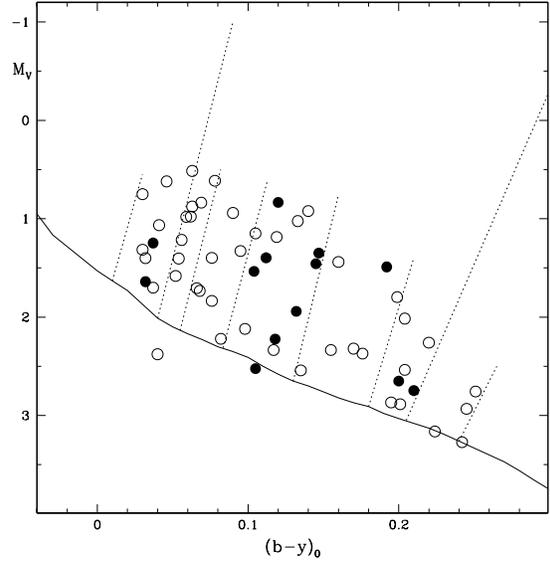}
\caption{Incidence of variability in our sample. The boxes are selected 
taking the borders roughly parallel to the blue and red borders (the
longest ones) of the instability strip. Filled circles: variable stars. 
Open circles: constant stars.}
\label{var} 
\end{figure}

Figure~\ref{var} shows the positions of already known
and new $\delta$ Sct stars located in the lower part 
of the instability strip. For sake of homogeneity, we considered only 
well--established variable and constant stars, i.e., the unclear cases
discussed in Sect. 4.2 are omitted. The magnitude limit $V\le$8.0 
implies that we observed as far as 310~pc in a cone with full aperture of
24\cc\, (galactic latitudes $b^{\rm II}$ between --12\cc\, and +12\cc) in the 
direction 
of the galactic centre (galactic longitudes $l^{\rm II}$ between +22\cc\,
and +45\cc), which in turn results in a maximum distance of 70~pc from
the galactic plane. 

\begin{figure} 
 \includegraphics[scale=0.40]{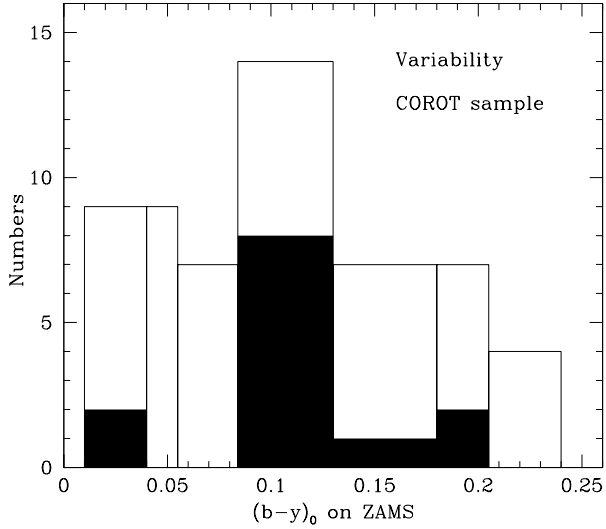}
\caption{Histograms of stars in the photometric boxes selected in
Fig.~\ref{var}. Note that the $(b-y)_0$ values are taken on the ZAMS. The
black areas indicate the number of pulsating stars, i.e., both $\delta$ 
Sct and $\gamma$ Dor variables.} 
\label{isto} 
\end{figure}

We arbitrarily divided the lower part of the instability strip in 7 boxes, 
 with margins roughly parallel to
the borders. The resulting histogram is shown in Fig.~\ref{isto}. Looking
at Figs.~\ref{var} and \ref{isto}, we can note that: 
\begin{enumerate} 
\item The variable stars are concentrated in the central box; 
\item Two stars, HD 170782 and the $\lambda$ Boo variable HD 188324 
are outside the instability strip on the blue edge; 
\item The two variables close to the red border are $\gamma$ Dor 
stars; 
\item Only one $\delta$ Sct star, HD 181555, is located near the ZAMS.
Most stars are more than 0.5 mag above it.
\end{enumerate}

Breger (2000) reports on the incidence of variability in the lower part of 
the instability strip by analysing stars spread all over the sky.
The incidence is around 50\% just in the 
central zone; when considering the whole instability strip the percentage
decreases to around 20\%.
In our sample we have 13 variables out of 57 stars (23\%). In the central
box we have 8 variables out of 14 (57\%). Of course, we have to stress
that percentages change as a function of the threshold of variability.
Both our and Breger's samples are limited to stars having an amplitude
larger than 3-4 mmag: therefore, it is no surprise to find the same 
percentages. 

\begin{figure} 
 \includegraphics[scale=0.40]{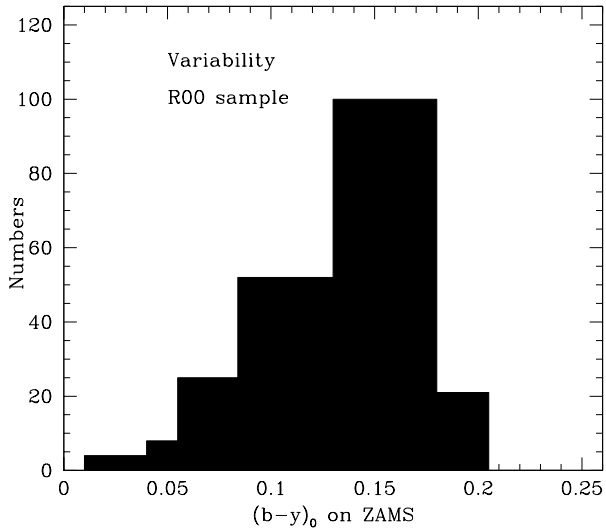} 
\caption{Histograms of $\delta$ Sct variables in the same photometric 
boxes considered in Fig.~\ref{var}, but considering the R00 sample 
(Rodr\I guez \& Breger 2001).}
\label{eloy} 
\end{figure}

However, the figures we obtained suggest that the incidence of
appreciable $\delta$ Sct photometric variability (i.e., the excitation of
modes with $\ell\le$3 resulting in light changes up to the mmag level and
more) is a function of the position in the CMD. To obtain more input on 
this point, we analysed the sample provided by Rodr\I guez \& Breger
(2001; R00 sample). We considered the 210 low--amplitude $\delta$ Sct 
stars located in the same part of the instability strip: we thus handled a 
group of stars similar to ours, except for the space distribution. 
 We obtained the histogram shown in Fig.~\ref{eloy}, 
where variability is well represented in the central box and is also more
common toward the red edge. As can be derived from the magnitude limit
and galactic coordinates reported above, the COROT sample is representative
of the disk population located in the solar 
neighbourhood. Comparing
 Figs. ~\ref{isto} and ~\ref{eloy}, the COROT sample appears slightly bluer than
the R00 sample. If real,
such a difference can originate only in the specific environment 
where COROT stars are located;
it is diluted when the sample loses in homogeneity, i.e., when it is extended
to the whole Galaxy. Therefore, the particular chemical composition
of solar neighbourhood  seems to enhance variability in hot stars.  Selection
effects are also plausible. Indeed, even if the photometric accuracy of our data
is quite good, our limited time sampling can miss the $\delta$ Sct
stars located closely to the red border, where the amplitudes are smaller.

We also note that HD 170782 re--opens the problem of variables 
beyond the blue border of the instability strip. Rodr\I guez \& Breger (2001) 
found a few candidates, but they suggest that the variability was
proposed on the basis of inaccurate photometry. 
In the case of HD 170782 the light variability is quite evident 
(Fig.~\ref{zoo}) and the position in the CMD too far from the blue border
to be ascribed to uncertainties in the $(b-y)_0$ photometry. Pamyatnykh (2000)
demonstrated that $\delta$ Sct pulsation should not occur for these stars in 
the classical picture. 
To have a solution, all these stars should be studied individually. For 
example, some of them may have higher helium abundance (in which case the
blue edge will be hotter), for some of them standard
photometric calibrations may not
apply owing to chemical peculiarities,  or
some other parameter may play a  subtle r\^ole.

\subsection{The rotational velocity of photometrically variable stars}

\begin{figure} 
 \includegraphics[scale=0.40]{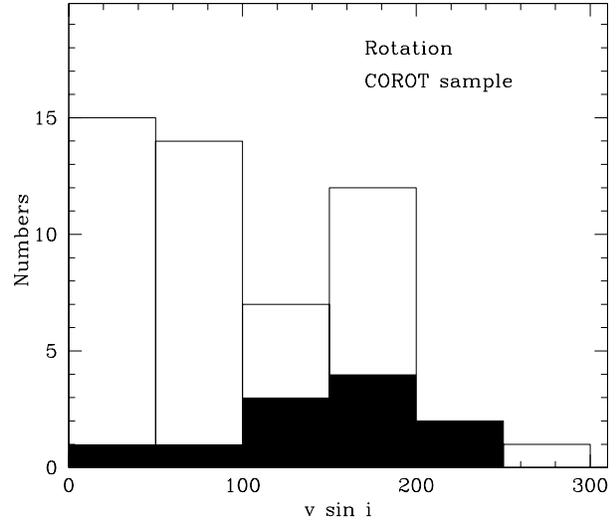} 
\caption{Histograms of $\delta$ Sct variables (black area) and constant 
stars (white area) as function of the measured $v\sin i$.}
\label{isto2}
\end{figure} 
\begin{figure} 
 \includegraphics[scale=0.40]{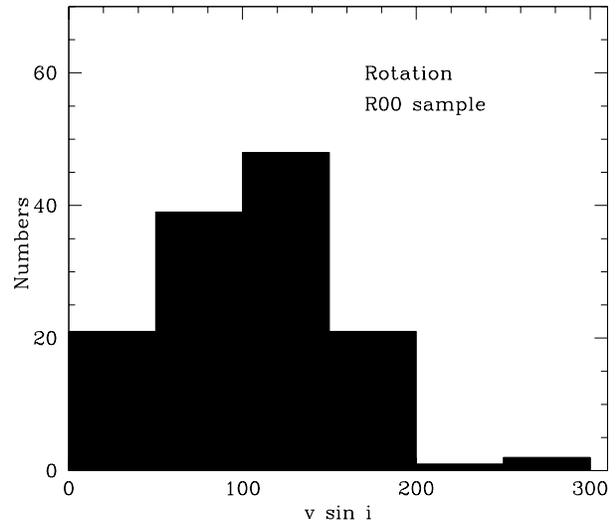}
\caption{Histograms of the $v\sin i$ values of $\delta$ Sct variables in 
the R00 sample (Rodr\I guez \& Breger 2001).}
\label{eloy2}
\end{figure}

There is a well--established relationship between photometric amplitude 
and $v\sin i$ values, i.e., high--amplitude pulsators are slow rotators 
(see Fig.~5 in Breger 2000). For amplitudes larger than 0.10 mag, $v\sin 
i$ is below 50~\kms. In our sample there is the remarkable 
exception of HD~177702 (see Fig.~\ref{zoo}, curve at bottom), which shows
a $v$ amplitude of 0.17 mag, which scales to about 0.12 mag in $V$ light,
and $v\sin i >$ 200~\kms. The star is far from the ZAMS and its
combination of high amplitude, fast rotation and evolved evolutionary 
state is unusual amongst the $\delta$ Sct stars. Further investigations 
are necessary to characterise HD~177702 both from a theoretical (evolution 
and rotation) and observational (high amplitude as a result of 
multiperiodicity?) point of view.

Let us investigate the effect of rotation on variability. 
We remind the reader that the break--up velocity is around 250~\kms\, for 
masses and radii typical for $\delta$ Sct variables. 
Considering the
11 $\delta$ Sct stars, 6 objects have $v\sin i >$150~~\kms, and 9 stars 
have $v\sin i >$100~\kms. Moreover, when considering all the stars having
$v\sin i >$150~~\kms, 9 are found to be photometrically constant, 6 are 
found to be variable. Figure~\ref{isto2} summarises what we found in our 
sample: it appears that light variability is concentrated in the 100--200 
~\kms\, range.
 When 
considering the R00 sample, the histogram of the available $v\sin i$ values
(132 stars out of 210 previously selected) shows 
the highest peak in the 100--150 ~\kms\, range (Fig.~\ref{eloy2}). 
The strong dependence on an unknown  parameter as the inclination angle 
$i$ smoothes the differences between the two samples, preventing physical
inferences. We can just suggest that a
rotational velocity of 100--150 ~\kms\, is quite common among $\delta$
Sct stars.

\begin{figure} 
\includegraphics[scale=0.40]{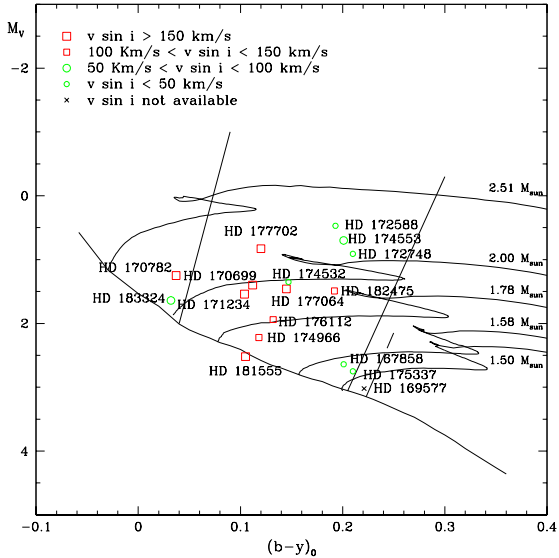} 
\caption{Hipparcos unreddened absolute visual magnitudes $M_V$ against
dereddened ($b-y$)$_0$ data for stars of interest in the Center direction. 
The theoretical tracks with $d_{\rm{over}}=0.2$ are shown.
All the variables are $\delta$ Sct stars, with the exception of
the $\gamma$ Dor stars HD 169577, HD 167858 and HD 175337. }
\label{hr}
\end{figure} 

\begin{figure} 
\includegraphics[scale=0.40]{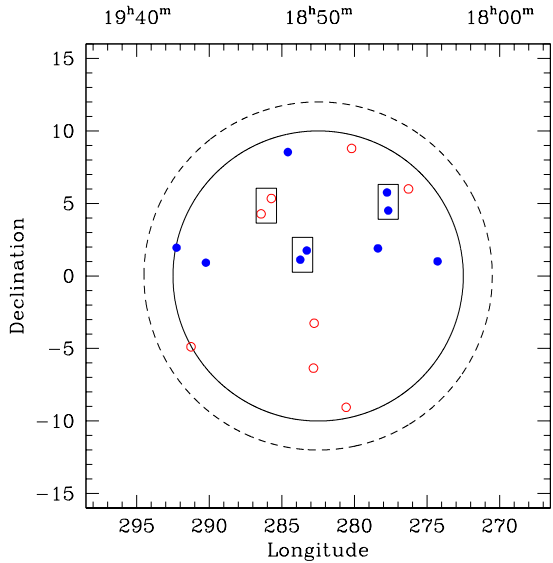}
\caption{The location of the stars of interest in the COROT accessible
field (solid and dashed lines, indicating different levels of stray light
contamination).
 Filled circles are high--priority stars, open circles are low--priority ones.
Three pairs of stars can be observed together in the Seismo 
CCD (whose field--of--view is indicated by the small rectangles).
 } 
\label{eyes} 
\end{figure}

\subsection{Our results in the COROT framework} 

By combining the results from Fig.~\ref{cmd} and Fig.~\ref{var}, we obtain 
Fig.~\ref{hr}. The evolved $\delta$ Sct stars HD 172748$\equiv\delta$ Sct, 
HD 172588, HD 174553 and the faint $\gamma$ Dor variable HD 169577 
($V$=8.64; see Sect.~6) are added for completeness. This figure illustrates the 
scientific output of the action undertaken to identify new potential COROT 
targets in the Center direction. We note that by observing a zone of 450 
deg$^2$ on or just above the galactic plane we discovered variables well 
mapping the lower part of the instability strip.

Figure~\ref{eyes} shows the positions of the stars of interest in the 
COROT field--of--view. The solid circle indicates the 10\cc\, cone, where
COROT will supply best results. The observing cone can be extended up to 
12\cc\, (dashed line) for objects having a special interest; above that 
limit stray light would cause a strong degradation in the photometric 
accuracy. Three pairs of stars are observable in the same field--of--view
of the Seismology CCD: HD 170782--HD 170699 (right), HD 174966--HD 175337
(middle) and HD 177702--HD 177064 (left). 

HD~174532 is the only $\delta$ Sct star showing a low $v\sin i$ value 
(32~\kms). None of the $\delta$ Sct stars close to 
ZAMS has a small $v\sin i$ (the lower limit is the 90~\kms\, observed for
the $\lambda$ Boo variable HD~183324). The search for smaller amplitude 
variability (less than 3 mmag) in slow rotators close to ZAMS (see 
Tab.~\ref{cst} for suitable targets) will be the goal of future dedicated
observations; it should also allow us to improve the percentages discussed 
above, as we will be able to infer how many variables we missed in the previous 
search. 

\section {The open cluster NGC 6633} 

The open cluster NGC~6633 is also located in the field accessible to
COROT.  A multisite campaign (Mart\I n, Alonso et al., in preparation)  
has been planned to search for variable stars: indeed, the cluster could
be a suitable target for COROT. The turn--off point is just above the
blue edge. With a distance modulus $m - M$=7.77, the members located in
the lower part of the instability strip are too faint (9.5$<V<$11.2) to be
COROT primary targets, but they can provide a further test for the
incidence of variability in a well--defined environment.

We established the membership by using the BDA database developed by
Mermilliod (1995), Sanders (1973) and Hiltner et al. (1958). A preliminary
analysis allowed us to firmly decide on the variability of six (five
$\delta$ Sct stars and one $\gamma$ Dor star) out of 30 members. In the
case of two $\delta$ Sct stars the membership is not conclusive; moreover,
the available photometry on additional six members does not allow to
choose between variability or constancy.  Therefore, we can propose a
value of 20\% (6 out of 30) for the incidence of variability, with small
fluctuations owing to the suspected stars and the uncertain memberships.
Such a result substantially confirms the value obtained for isolated
stars.

Cluster variables of other classes have been also found (among them, two
Ap stars and one eclipsing binary brighter than $V$=9.5). Other pulsating
stars are not included in the above statistic as they are considered
non--members: the previously known $\gamma$ Dor star HD~169577 ($V$=8.64)  
(Mart\I n \& Rodr\I guez 2002), the two new multiperiodic $\delta$ Sct
stars HD 169597 ($V$=9.13) and BD+06\cc3737 ($V$=9.15). Therefore, in
Fig.~\ref{eyes}, the open circle close to the pair at the right marks not
only the position of HD 169577, but also the position of all the bright
variables described here. By adding the pair itself (the two bright
$\delta$ Sct stars HD 170782 and HD 170699), that corner of sky looks very
promising for a COROT investigation.

\section{Conclusions} 

The search for new targets in the COROT accessible field resulted in an 
exercise of variability detection in stars located in the solar
neighbourhood and belonging to the disk population, i.e., in a 
well--defined galactic region. 
$\delta$ Sct stars are commonly found, the
incidence of this variability is around 25\% of the stars located in the 
lower part of the instability strip. A similar percentage is observed 
in the open cluster NGC~6633. The typical pulsator is a
multiperiodic one, a little evolved, in the middle part of the strip or 
slightly cooler. Unfortunately, even if slow rotators are common in the 
Galaxy, we generally found fast ones ($v\sin i >$ 100 \kms) and just 
one with a really low $v\sin i$, HD 174532. A spectroscopic analysis of the
line profile variations will be 
helpful to estimate the inclination angle $i$ of this promising target. 

In the selected sample 18 stars were located close to or beyond the blue
border of the instability strip: in addition to the previously known
$\lambda$ Boo star HD 183324, we also found one new $\delta$ Sct star. The
observational onset of pulsational instability in this region needs
further attention in the future, maybe on the basis of star--by--star
investigations.

The results we obtained met the scientific goals of the COROT mission,
ensuring a wide choice between $p$--mode pulsators; in general, $\delta$ 
Sct stars are confirmed to be suitable targets for any asteroseismic
space mission, taking into account elemental diffusion and richness of
excited modes. Stars appearing as a challenge to theory (HD 170782) or
unusual (HD 177702) were also discovered. 

\begin{acknowledgements} 
This research has made use of the {\sc simbad} database, operating at 
CDS, Strasbourg, France. The authors wish to thank A.~Pamyatnykh
and the referee, F.~Pijpers, for useful comments, and  
the STARE team for the attribution of observing time   
to the COROT project.
SM acknowledges financial support from a European Union Marie Curie
Fellowship, under contract HPMF-CT-2001-01146. RG 
acknowledges financial support from the program ESP2001-4528-PE. 
TL, NN and WWW was supported by the Austrian Fonds zur F\"orderung der
wissenschaftliche Forschung (P14984) and the BM:WUK (project COROT).
PJA acknowledges financial support at the Instituto de Astrof\'{\i}sica 
de Andaluc\'{\i}a-CSIC by an I3P contract (I3P-PC2001-1) funded by the 
European Social Fund.
\end{acknowledgements} 
 
\end{document}